\documentclass[%
 reprint,
superscriptaddress,
 amsmath,amssymb,
 aps,
twocolumn,
prb,
]{revtex4}
\usepackage{graphicx}
\usepackage{times}
\usepackage{dcolumn}
\usepackage{bm}
\usepackage{colordvi}
\usepackage{color}
\usepackage{epsfig}

\begin{document}

\title{Iterative Entanglement Distillation: Approaching full Elimination of Decoherence}

\author{Boris Hage}
\affiliation{Albert-Einstein-Institut, Max-Planck-Institut f\"ur Gravitationsphysik and Leibniz Universit\"at Hannover, Callinstr. 38, 30167 Hannover, Germany}

\author{Aiko Samblowski}
\affiliation{Albert-Einstein-Institut, Max-Planck-Institut f\"ur Gravitationsphysik and Leibniz Universit\"at Hannover, Callinstr. 38, 30167 Hannover, Germany}

\author{James DiGuglielmo}
\affiliation{Albert-Einstein-Institut, Max-Planck-Institut f\"ur Gravitationsphysik and Leibniz Universit\"at Hannover, Callinstr. 38, 30167 Hannover, Germany}

\author{Jarom\'{i}r Fiur\'{a}\v{s}ek}
\affiliation{Department of Optics, Palack\'y University, {17. listopadu 12, 77146 Olomouc,}
Czech Republic}

\author{Roman Schnabel}
\affiliation{Albert-Einstein-Institut, Max-Planck-Institut f\"ur Gravitationsphysik and Leibniz Universit\"at Hannover, Callinstr. 38, 30167 Hannover, Germany}

\date{\today}

\begin{abstract}
The distribution and processing of quantum entanglement form the basis of quantum communication and quantum computing. The realization of the two is difficult because quantum information inherently has a high susceptibility to decoherence, i.e. to uncontrollable information loss to the environment. For entanglement distribution, a proposed solution to this problem is capable of fully eliminating decoherence; namely iterative entanglement distillation. This approach builds on a large number of distillation steps each of which extracts a number of weakly decohered entangled states from a larger number of strongly decohered states. Here, for the first time, we experimentally demonstrate iterative distillation of entanglement. Already distilled entangled states were further improved in a second distillation step and also made available for subsequent steps.Our experiment displays the realization of the building blocks required for an entanglement distillation scheme that can fully eliminate decoherence.
\end{abstract}

\maketitle

\begin{figure*}[t]
\includegraphics[width=\textwidth]{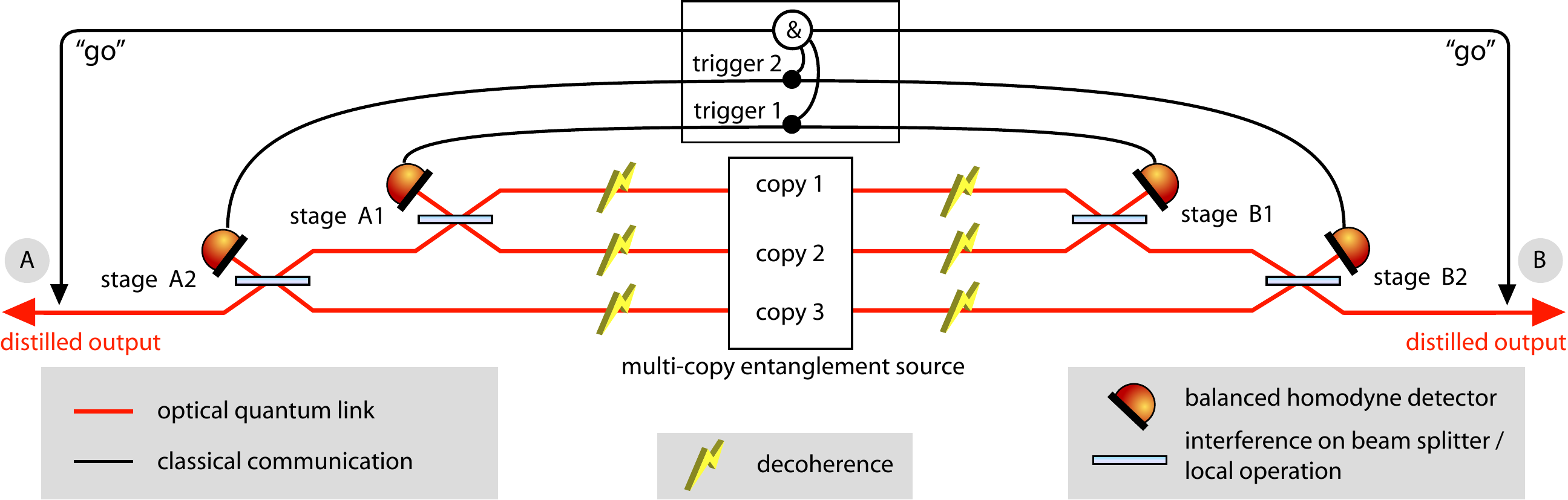}
\caption{(Color online) Schematic of the experiment. The Iterative three-copy distillation was demonstrated in the following way. When the entangled states of light left the source they decohered due to phase fluctuations from coupling to the environment. Before they completely lost their entanglement a first two-copy distillation protocol was realized selecting states with low decoherence and strong entanglement by means of local measurements and classical communication. From the already distilled states another subsequent two-copy distillation step further counteracted decoherence. For an infinite number of two-copy distillation steps decoherence can be fully eliminated.
}
\label{setup}
\end{figure*}

If two or more subsystems are in an entangled state the physical realities of these subsystems can not be seen as independent from each other~\cite{EPR35}, although they do not interact and may be separated by an arbitrarily large distance. For maximally entangled states, a measurement performed on an individual subsystem (a so-called local measurement) provides a completely random result that, surprisingly, exhibits a correlation, stronger than any classical correlation, with the also locally-random measurement result from another subsystem.  This remarkable phenomenon {is exploited in quantum communication~\cite{Alber01} and quantum computing~\cite{Nielsen2000}. Examples are} secure communication via quantum cryptography~\cite{Ekert91,Gisin02} or quantum teleportation~\cite{Zeilinger97,Furusawa98,Bowen03}.

Entanglement can be efficiently distributed by using entangled pairs of photons or entangled pairs of light modes. The transmission channels, be it optical fibres or free space as envisioned in satellite-based quantum cryptography~\cite{Aspelmeyer03}, are however inevitably lossy, and the coupling to the {environment} leads to decoherence and to the degradation of the initial entanglement. The decoherence thus imposes a fundamental limit on communication distance, information processing time and complexity beyond which the entanglement is completely destroyed and the advantages of using quantum -- rather than classical -- systems are lost.

It was therefore one of the great discoveries in the early years of quantum information science that the deleterious effect of noisy channels on entanglement distribution can be counteracted \emph{a-posteriori} by operations performed locally on each part of the shared entangled state and classical {communication~\cite{Bennett96,Deutsch96,Bennett96concentration,Eisert04}.} The conceptually simplest approach is the single-copy entanglement distillation whereby a local quantum filter is applied to one part of the shared entangled state~\cite{Bennett96concentration,Kwiat01,Dong08,Takahashi09}. Successful filtration results in a probabilistic increase of entanglement of the shared state. However, this method is intrinsically limited in its ability to overcome decoherence and enhance the entanglement and purify the state~\cite{Linden98}. A more sophisticated approach is a distillation protocol that uses the local interference of two copies of the entangled {states~\cite{Eisert04,Fiurasek07}.} Such entanglement distillation schemes were demonstrated recently~\cite{Pan03,Zhao03,Hage08} but a single application of these protocols does not allow to completely eliminate decoherence or the transmission over an arbitrary distance. A solution to the problem of decoherence is provided by the \textit{iterative} distillation schemes~\cite{Bennett96,Eisert04,Marek07}. Such schemes (i) involve the distillation of input states that have already been distilled in a previous step, and (ii) enable the application of  subsequent steps. The entanglement of the distilled state increases with each successful iteration, and under certain conditions the protocol asymptotically converges to a maximally entangled pure state.

Here we report on the first experimental demonstration of iterative entanglement distillation. We implemented a two-step distillation protocol that used three decohered copies of an entangled state shared between two parties A and B, as shown in Fig.~\ref{setup}. The first distillation step used two decohered copies of the shared entangled state. At both locations the local parts of the two states were superimposed on a balanced beam splitter with one output port at each site  detected by a balanced homodyne detector. A comparison of the measurement results via classical communication yielded a probabilistic signal for the success of distillation. The protocol provided one output copy with increased shared entanglement and partially eliminated decoherence\cite{Hage08}. The second two-copy distillation step then used the already distilled state and a third decohered copy. Again local measurements and classical communication were used to prepare an iteratively distilled state with even less decoherence and even higher entanglement. After successful two-step distillation the state was available for subsequent distillation steps, or, alternatively e.g.~for a quantum teleportation protocol. The output state was characterized by means of a full two mode quantum state tomography~\cite{Leonhardt96,Raymer96}.

Our experiment used three copies of continuous-variable entangled pairs of continuous-wave laser beams. Each entangled pair was generated by splitting a squeezed laser mode on a balanced beam splitter. The squeezed states of light were generated in optical parametric amplifiers (OPAs), which were constructed from second order non-linear crystals ($\textrm{MgO:LiNbO}_3$) with type-I phase matching inside a degenerate doubly resonant cavity~\cite{CVHFLDS05}. The OPA process was pumped with a frequency doubled laser beam at 532\,nm originating from a monolithic solid state laser (Nd:YAG) operating at $1064\,\textrm{nm}$. All three OPA outputs showed about 5\,dB of squeezing and 9\,dB of anti-squeezing at modulation frequencies ranging from 5\,MHz to 15\,MHz.

All the three entangled pairs generated were distributed to two locations A and B and were intentionally exposed to independent random phase fluctuations which led to decoherence and to a degradation of the entanglement and state purity~\cite{Franzen06}. In our experiment the phases of each of the six light fields involved were individually diffused by piezo actuated phase shifters driven by Gaussian quasi-random voltages that were generated by a PC sound card. Random phase fluctuations are a rather natural decoherence source, which produces non-Gaussian statistics of the states. This non-Gaussian property of phase noise allowed us to use Gaussian operations within our iterative distillation steps\cite{Eisert04,Fiurasek07}. Note that for a purely Gaussian framework a no-go theorem for distillation applies~\cite{Eisert02,Giedke02}. \begin{figure*}[t]
\includegraphics[width=\textwidth]{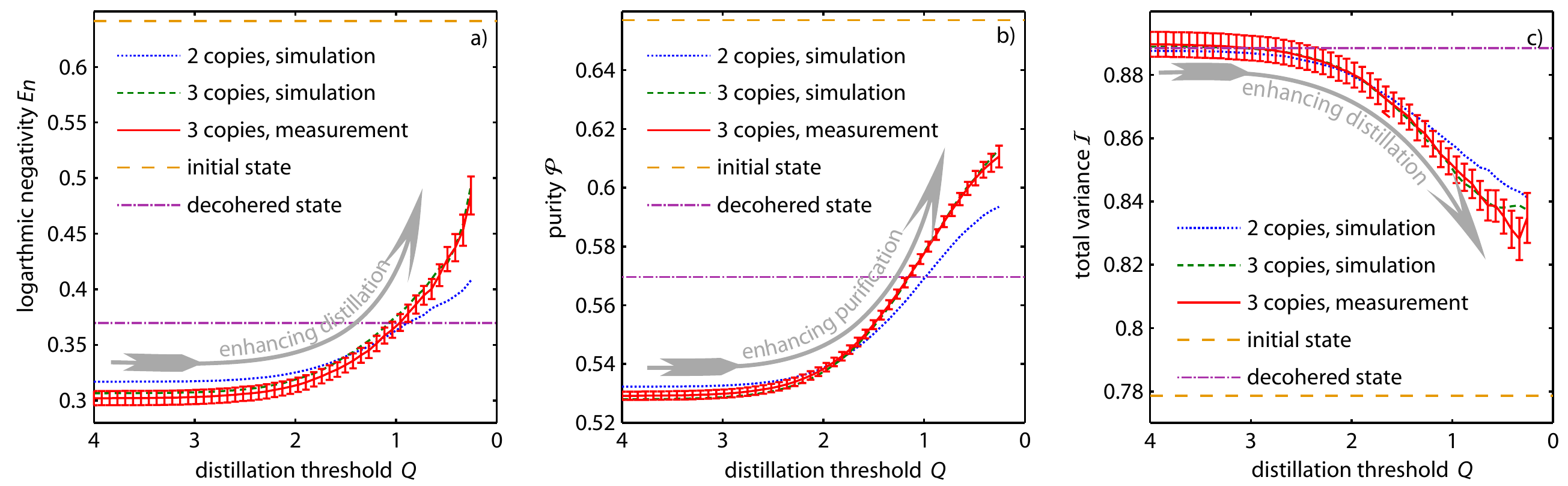}
\caption{\label{MainResult}(Color online)  Experimental data are given as solid line including error bars. Plots (a)-(c) show the logarithmic negativity, purity, and the total variance of the distilled states versus trigger threshold applied to both distillation stages. The short dashed lines show the result of Monte Carlo simulations with the exact parameters of the experiment. Note, that the error bars of the simulation (not shown) have the same magnitude as those of the measurement data, because both are dominated by the statistics of the sample number of $3\!\cdot\!10^5$ for each of $10^2$ tomography slices. The dottet lines represent the numerical simulation for the corresponding single-step two-copy protocol assuming exactly the same experimental parameters. The long dashed lines represent the values before the decoherence, i.e. without any phase noise. The dot-dash lines characterize the decohered input states before the distillation stages. All three quantities are improved by the distillation beyond their respective values for the input states. The iterative distillation outperforms the corresponding single-step protocol.}
\label{MainResults}
\end{figure*}
The local operations and quantum measurements in our iterative distillation experiment involved the interference of sixteen laser beams on eight beam splitters. Four beam splitters were required to pairwise interfere the distributed parts of the three entangled copies for the two distillation steps. Those in the first stage were balanced while those in the second stage provided a 2:1 power \textit{transmittance:reflectance} ratio. Another four beam splitters were integral parts of the four balanced homodyne detectors (BHDs) which are shown in Fig.~\ref{setup}. In the BHDs the beam splitters were used to generate an interference signal with a local oscillator beam which then provided the local quantum measurements. Another two beam splitters were used in two more BHDs that were placed in the output ports of our setup (not shown in Fig.~\ref{setup}) in order to independently verify and characterize our distillation protocol by means of a full quantum tomography on the distilled outputs at A and B. The fringe contrasts achieved at the BHDs and at the distillation stages were between 97\% and 99\%.
An important aspect of our experiment was the simultaneous phase control of the two-times ten laser beam inputs to the beam splitters mentioned. In order to generate error signals for the in-phase interference at the distillation stages small fractions (3\%) of the beams were tapped in front of the balanced homodyne detectors in the distillation stages.

The \textit{"go"}-signal for successful iterative distillation in our experiment required (positive) trigger signals from both (logical AND) of the distillation stages. In each stage {a trigger signal} was generated from two BHDs photo-current outputs: A1 and B1 or A2 and B2, respectively (Fig.~\ref{setup}). In both BHDs of a stage the phases of the local oscillators were controlled such that the quantum-correlated field quadratures of the signal beams were measured. The difference of the photo-currents provided information about how likely it was to have better than on average entanglement on the second un-measured output of the preceding distillation beam splitter, as shown in Refs.\cite{Fiurasek07,Hage08}. The difference photo-currents of the two distillation stages were amplified by home-made electronics, converted into voltages and electronically mixed with an electronic local oscillator at the radio-frequency of 7\,MHz which in our case corresponded to the regime of highest initial entanglement. The voltage signals were then anti-alias filtered with a bandwidth of 400\,kHz and synchronously sampled with 1\,MHz. The result were two time series of voltage values fluctuating around zero. If an absolute voltage value was below a fixed but variable threshold value of our choice, the trigger for successful distillation was positive. If both triggers were positive, successful iterative distillation was indicated ("\textit{go}"). The lower the thresholds were, the stronger the distillation and purification effect was and the lower the total distillation yield became.

In order to completely characterize the distillation protocol we performed a full tomographic reconstruction of the iteratively distilled states at the output ports of the experiment for several different values of the trigger thresholds. The elements of two-mode density matrix in the Fock state basis $\rho_{nklm}=\langle nk|\hat\rho|lm\rangle$ were obtained by averaging the appropriate pattern functions $S_{nk}^{lm}$ over the recorded homodyne data~\cite{Leonhardt96,Raymer96}. Since the magnitude of the matrix elements decays rapidly for higher photon numbers, the reconstruction was truncated for $\left\{n,k,l,m\right\}\geq5$. Fig.~\ref{MainResult} presents the result of our work derived from our tomography data and illustrates that our iterative distillation protocol increases the entanglement and purity, and outperforms the corresponding single-step distillation protocol. We plotted the logarithmic negativity (a), the purity (b), and the total variance (c), respectively, versus the trigger threshold value. The latter was kept equal in both stages since our explicit data analysis showed that the best performance is generally achieved if the threshold values of both stages are the same. Our measurement data is given as a solid line including their error bars dominated by sample statistics. A precise Monte Carlo simulation of our experimental setting is given as a short dashed line. Our model reproduces the measurement very well. Since the same sample size was used the size of the statistical fluctuations also corresponded to the measurement error bars. The comparison with the corresponding single-stage distillation protocol is again given as a precise numerical simulation (dotted line). Note that a comparison with experimental data from a single-stage distillation protocol would be less accurate since a complete re-arrangement of the experiment was necessary. The long dashed lines charactrize the initially prepared states without phase noise. The dot-dash lines display the impact of the intentional phase diffusion. Note, the mixing of the input copies in the distillation stages further lowered the logarithmic negativity and the purity (see the left end of the individual figures).  

The logarithmic negativity $E_n$ as given in Fig.~\ref{MainResult}a) is a computable {monotone} measure of the entanglement strength~\cite{Vidal02}.  {Presence of entanglement is certified if $E_n>0$, and, furthermore, the higher the value of $E_n$ the stronger the entanglement is.} Fig.~\ref{MainResult}a) shows that the entanglement strength of the distilled states increased with lower (more strict) threshold values and exceeded the value of the input states (dot-dash line). This break even point corresponded to a probability of success of more than 70\%/50\% for the two/three copies protocols.  Indeed, the iterative protocols yielded a faster increase than in the corresponding single-stage protocol. When changing the sample number $N$ in our Monte Carlo simulations we found that the logarithmic negativity slightly decreased with increasing number of samples~\cite{footnote}. This effect was also observed in Ref.~\cite{Takahashi09}. Since the results shown in Fig.~\ref{MainResult}a) were all obtained for the same number of samples (measured and simulated) a fair comparison is guaranteed. Fig.~\ref{MainResult}b) demonstrates that also the purity of the distilled state, as given by $\mathcal{P}=\mathrm{Tr}(\rho^2)$, did increase beyond the input states, again taking advantage from the iterative two-stage protocol for lower threshold values.
In Fig.~\ref{MainResult}c) we used another measure of entanglement which is relevant for entanglement with a Gaussian statistics and for downstream applications within a purely Gaussian setting, such as the teleportation of Gaussian states~\cite{Furusawa98,Bowen03}. Plot c) shows the corresponding states' \textit{total variance} {$\mathcal{I}$ whose definition is based upon the variances of the difference and sum of the amplitude quadrature measurement results and the phase quadrature measurement results at parties $A$ and $B$ ($X_A$, $X_B$, $P_A$, $P_B$), respectively: $\mathcal{I}=Var(X_A - X_B) + Var(P_A +  P_B)$. In this work we normalized the quadrature variance of an individual mode in its ground state to 1/4. Then, according to Ref.~\cite{Duan00}, the state is entangled if $\mathcal{I}<1$. Note that for this entanglement measure a \textit{smaller} value of $\mathcal{I}$ corresponds to stronger (Gaussian) entanglement.} 
\begin{figure}[t]
\includegraphics[width=0.75\columnwidth]{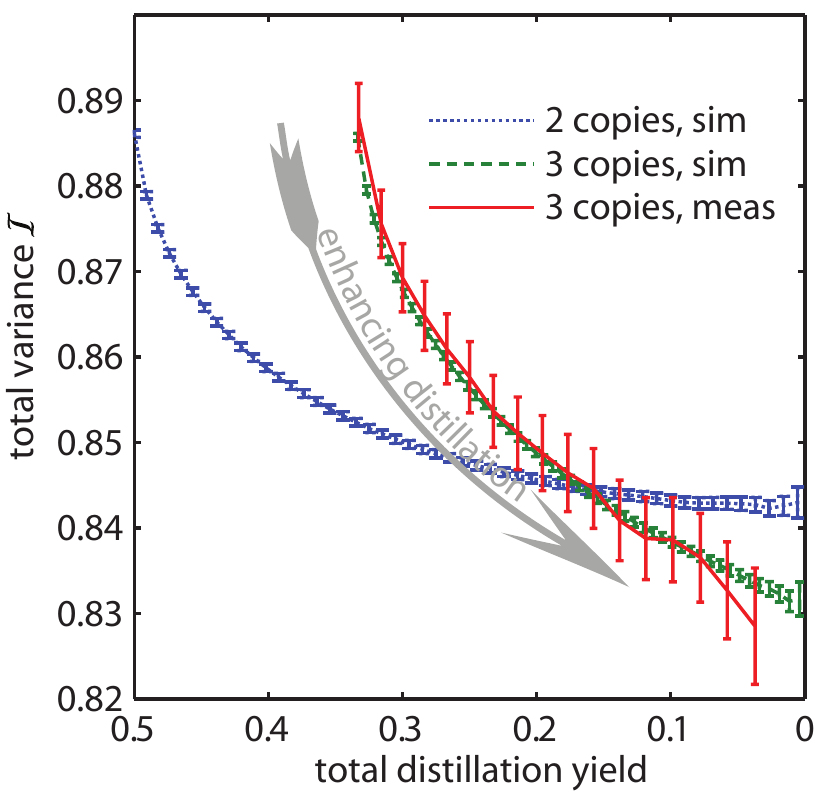}
\caption{\label{yield}(Color online) Our three-copy iterative distillation is superior even under the more strict condition that not the trigger threshold(s) but the total distillation yield is kept constant. Here, the simulations used a 300 times higher sample number providing smaller error bars.
}
\end{figure}

An interesting question is how the improved performance due to the second distillation stage effects the overall rate of the distillation yield. Remarkably, Fig.~\ref{yield} shows that in the relevant regime of a significant distillation effect the iterative scheme is superior even for a fixed requested total distillation yield. As an example, consider 3000 initially distributed decohered entangled states and a total distillation yield of 10\%. With the single-stage protocol the distillate is the result from 1500 two-copy distillations and contains 300 states with a total variance of $\mathcal{I}=0.843$.
With our iterative two-stage protocol the distillate is the result of 1000 three-copy distillations and also contains 300 states but with stronger entanglement corresponding to a total variance of just $\mathcal{I}=0.838$. 
Note that in this plot the numerical simulations (dashed, dotted) used {300 times higher sample numbers (3$\times10^6$)} which resulted in statistical error bars considerably smaller than those of the measurement data (solid). The simulated curves clearly suggest that our two-stage iterative protocol improved the entanglement into a regime not accessible for a single stage protocol.

In our experiment we quantitatively and qualitatively analyzed the entanglement shared between two separated locations A and B when an iterative distillation protocol was applied in order to counteract decoherence. We have successfully shown that an already distilled state can be further distilled when another decohered copy of shared entanglement was integrated. {Complete evidence was provided by the first realization of a full two-mode continuous-variable tomographic reconstruction of the entangled states}. 
A remarkable result of our experimental and theoretical analysis is that our entanglement distillation protocol, though iterative, provides a surprisingly high efficiency. The protocol provides a significant distillation and purification effect combined with a high total distillation yield of the order of 10\%. We also emphasize that our iterative distillation protocol does not depend on the characterization at the output ports and that the distilled states can be used in any downstream quantum protocol. The distilled states can therefore also be used as the input for further distillation stages. In combination with a de-Gaussification protocol as recently used in Ref.~\cite{Takahashi09} our iterative distillation experiment can be used in order to counteract also optical loss and not only phase diffusion as considered here. We therefore experimentally realized the necessary building blocks for a feasible full elimination of decoherence. Our result might stimulate applications in quantum communication since it was experimentally demonstrated how the main handicap, the existence of decoherence, can be overcome. 

J.F. acknowledges the financial support of
the Future and Emerging Technologies (FET) programme within the
Seventh Framework Programme for Research of the European Commission,
under the FET-Open grant agreement COMPAS, number 212008 and from
MSMT under projects Center of Modern Optics (LC06007) and
Measurement and Information in Optics (MSM6198959213). R.S.
acknowledges financial support from the Deutsche
Forschungsgemeinschaft (DFG), Project No. SCHN 757/2-1 and the Centre for Quantum Engineering and
Space-Time Research QUEST.


\begin{thebibliography}{}
\bibitem{footnote}
The dependence of $E_n$ on $N$ could be described by an empirical relation
  $E_n=a+b/N^c$, with $c \approx 0.6$, which is consistent with Ref.{
  }\cite{Takahashi09}.
\end{thebibliography}

\end{document}